# Peer to Peer Networks Management Survey

**Mourad AMAD[1], Ahmed MEDDAHI[2] and Djamil AÏSSANI[3]**

[1,3] Laboratory LAMOS, University of Bejaia, Algeria
*mourad.amad@univ-bejaia.dz, lamos_bejaia@hotmail.com*

[2] Institut Telecom/Telecom Lille 1, France
*ahmed.meddahi@telecom-lille1.eu*

**Abstract**

Peer-to-Peer systems are based on the concept of resources localization and mutualisation in dynamic context. In specific environment such as mobile networks, characterized by high variability and dynamicity of network conditions and performances, where nodes can join and leave the network dynamically, resources reliability and availability constitute a critical issue. The resource discovery problem arises in the context of peer to peer (P2P) networks, where at any point of time a peer may be placed at or removed from any location over a general purpose network. Locating a resource or service efficiently is one of the most important issues related to peer-to-peer networks. The objective of a search mechanism is to successfully locate resources while incurring low overhead and low delay. This paper presents a survey on P2P networks management: classification, applications, platforms, simulators and security.

## 1. Introduction

Peer-to-Peer (P2P) systems are distributed systems without (*or with a minimal*) centralized control or hierarchical organization, where each node is equivalent in term of functionality. P2P refers to a class of systems and applications that employ distributed resources to perform a critical function such as resources localization in a decentralized manner. The main challenge in P2P computing is to design and implement a robust distributed system composed of distributed and heterogeneous peer nodes, located in unrelated administrative domains. In a typical P2P system, the participants can be "domestic" or "enterprise" terminals connected to the Internet.

Peer-to-Peer computing is a very controversial topic. Many experts believe that there is not much new in P2P. There are several definitions of P2P systems that are being used by the P2P community. As defined in **[48]**, "P2P allows file sharing or computer resources and services by direct exchange between systems", or "allows the use of devices on the Internet periphery in a non client capacity". Also,"it could be defined through three key requirements: **a)** they have an operational computer of server quality, **b)** they have a DNS independent addressing system" and **c)** they are able to scope with variable connectivity. Also, as defined in **[61]**: P2P is a class of applications that takes advantage of resources-storage, cycle, content, human presence-availability at the edges of Internet. Because accessing to these decentralized resources means operating in environment with unstable connectivity and unpredictable IP addresses. P2P nodes must operate outside the DNS system and have significant or total autonomy from central servers **[48].**

In this paper, we give a generalized state of art on P2P networks; we present different classifications with representative examples, P2P platforms and applications, major security problems and P2P simulators with analysis and discussion.

## 2. Peer to Peer Network Architectures

P2P systems implement a virtual overlay network over the underlying physical network as described on Figure 1. There are several proposed architectures. In this section, we give some descriptions of the most important P2P protocols with a classification based on the underlying architecture. Based on this concept, we can classify them into two categories: structured and unstructured systems.

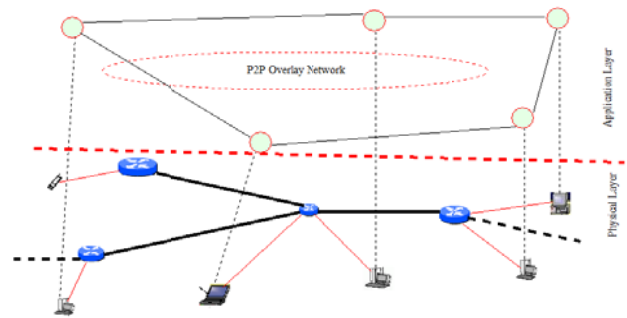

**Figure 1:** P2P Overlay Network

## 2.1 Unstructured Peer to Peer Networks

Unstructured P2P systems refer to P2P systems with no restriction on data placement in the overlay topology. Some P2P systems also provide the query functionality to locate the files by keyword search. The most significant difference of unstructured P2P systems to the traditional client/server architecture is the high availability of files and network capacity among the peers. Replicated copies of popular files are shared among peers. Instead of downloading from the central server, peers can download the files from other peers inside the network. Obviously, the total network bandwidth for popular file transmissions is undoubtedly greater than the amount most central server systems able to provide. In this sub section, we give functional principles of some unstructured Peer to Peer architectures.

### 2.1.1 Example illustrations

Napster [30] is known as music exchange system. Nodes login to a server and send a list of files that can offer, then issue queries to the server to find which other nodes hold their desired files, and finally download the desired objects directly from the object home. In BitTorrent [24], a peer that wishes to download a certain file joins a group of about 50 other peers that either upload or download this same file. This group is called the **swarm**. The file is split into many small chunks, and these chunks are exchanged by the peers in the swarm. The advantage of this is that peers that download a chunk can upload it to the other peers that do not yet have this chunk, so all chunks of the file are available from multiple peers. This is known as the barter mechanism. Gnutella [2] is a decentralized protocol for distributed search in a flat topology of peers (*servents*). In Gnutella, a querying peer sends a query to all of its neighbour peers, who in turn send the query to all of neighbour, and this spreading broadcast continues until the query reaches a peer that has a file that matches the query, or until a certain predefined maximal number of forwards are reached. If a peer is reached, it sends back a reply containing its address, the size of the file, speed of transfer, etc. The reply traverses the same path as the query but in a reverse order back to the querying peer. In this wary, query is propagated $N^p$ other peers (*where N is the number of neighbour peers and p is the maximum number of forwards 'TT' of the query*). This passing of messages generates much traffic in the networks often leading to congestion and slow response. Freenet [29] is a purely decentralized loosely structured system. It is essentially pools unused disk space in peer computers to create collaborative virtual file system providing file security and publisher anonymity. Freenet provides file-storage service rather than file-sharing service unlike Gnutella. In ECSP [12], peers are grouped into clusters according to their topological proximity, and super-peers are selected from regular peers to act as cluster leaders and service providers. These super-peers are also connected to each other, forming a backbone overlay network operating as a distinct application. The overlay is managed using an application level broadcasting. HPPC [14] (*Hierarchical Projection Pursuit Clustering*) is a clustering architecture; a cluster is characterized by regions of high density separated by regions that are sparse. DV-Flood is [6] organized as a clusters connected through gateways (*super nodes*). Each cluster is represented by a leader using an election/selection algorithm. DV-Flood uses flooding technique for resource retrieval and localization in inter-clusters and intra-clusters. In the first case the flooding is limited by a parameter *V* and in the second way, it is limited by a parameter *V*. The leaders are used only for routing acceleration, not for data replication or storage.

### 2.1.2 Analysis and Discussion

Unstructured P2P systems can support partial keyword search. These systems depend on blind search techniques, such as flooding and random walk. Hence, the generated volume of query traffic does not scale with the growth in network size. Many research activities are aimed towards improving the routing performance of unstructured P2P systems by adopting hint-based routing strategies. Peers learn from the results of previous routing decisions, and bias future query routing based on this knowledge. Unstructured P2P networks offer a number of important advantages: **(1)** an unstructured network imposes very small demands on individual nodes, and more specifically it allows nodes to join or leave the network without significantly affecting the system performance. **(2)** Unstructured networks are appropriate for content-based retrieval (*e.g., keyword searches*) as opposed to object identifier location of structured overlays. **(3)** Finally, unstructured networks can easily accommodate nodes of varying power. Consequently, they scale to large sizes and they offer more robust performance in the presence of node failures and connection unreliability. According to **[27]**, if scalability concerns were removed from unstructured P2P systems, they might be the preferred choice for file-sharing and other applications where the following assumptions hold: **(1)** Keyword searching is the common operation, **(2)** most content is typically replicated at a fair fraction of participating sites and **(3)** the node population is highly transient. Table 1 presents some complexities of unstructured P2P networks.

**Tab 1.** Complexities of some unstructured P2P systems

| Architecture | Space complexity | Cost lookup |
|---|---|---|
| Napster | *n* | *O(1)* |
| Gnutella | *O(n)* | *O(n)* |
| Freenet | *Hops to Leave* | *Hops to Leave* |
| DV-Flood | *O(D*V)* | *O(D*V)* |

## 2.2 Structured Peer to Peer Networks

Structured P2P networks have emerged mainly in an attempt to address the scalability issues that unstructured systems are faced with. The random search methods adopted by unstructured systems seem to be inherently not scalable [27], and structured systems were proposed, in which the overlay network topology is tightly controlled and files (*or pointers to them*) are placed at precisely specified locations. These systems provide a mapping between the file identifier and location, in the form of a distributed routing table, so that queries can be efficiently routed to the node with the desired file. Structured P2P systems are also referred to as Distributed Hash Tables (DHTs). The evolution of research in DHTs was motivated by the poor scaling properties of unstructured P2P systems.

Thus, a key contribution of DHTs is extreme scalability. As a result, DHTs exhibit some very unique properties that traditional P2P systems lack [49]. Generally, P2P overlay network is characterized by the decisions made on the following six key design aspects [54]:

**1)** Choice of an identifier space, **2)** Mapping of resources and peers to the identifier space, **3)** Management of the identifier space by the peers, **4)** Graph embedding (*structure of the logical network*), **5)** Routing strategy and **6)** Maintenance strategy.

In this sub section, we give functional principles of some structured P2P architectures.

### 2.2.1 Example illustrations

In Plaxton [31], each node or machine can take on the role of servers (*where objects are stored*), routers (*which forward messages*) and clients (*origins of requests*). Objects and nodes have names independent of their location and semantic properties, in the form of random fixed length bits sequences represented by a common base. Object location in plaxton works as follows: **1)** a servers S1 publishes that it has an objects O1 by routing a message to the root node of O1. The published process consists of sending toward the root node a message, which contains a mapping *<object-id, server-id>*. **2)** During object location, a query message destined for object O1 is initially routed towards O1's root. **3)** At each step, if the message encounters a node that contains the location mapping for O1, it is immediately redirected to the server containing the object O1. Otherwise, the message is forward one step closer to the root. If the message reaches the root, it is guaranteed to find a mapping for the location of O1. Chord [5] is a decentralized peer to peer lookup service that stores key/value pairs for distributed data items. Given a key, the node responsible for storing the key's value can be determined using a hash function that assigns an identifier to each node and to each key. Each key $k$ is stored on the first node whose identifier *id* is equal or follows $k$ in the identifier space. DKS [39] stands for Distributed k-ary Search and it was designed after perceiving that many DHT systems are instances of a form of k-ary search. A query arriving at a node is forwarded to the first node in the interval to which the *id* of the node belongs. Therefore, a lookup is resolved in $log_k(N)$ hops. CAN [3] is a distributed and structured P2P lookup services, each key will be evenly hashed into a point of d-dimensional space, as its identifier, when a node joins, it will randomly select a point of d-dimensional space. Then, it will be responsible for half of regions this point belongs to, and holds all keys who's IDs belongs to this region. Each node will keep its neighbour node ID locally, and routing is then performed by forwarding request to the regions closest to the position of the key. The expected search length is $O(d\sqrt[d]{N})$ and state information kept locally is $O(d)$. FCAN [32] is based on CAN and propose a kind of search scheme in structure P2P that supports semantic based query. The construction of P2P overlay should ensure that the organization of peers in P2P system and the placement of the data objects are consistent with the semantic space that they belong to. ABC [7] is also called Alpha-Beta cluster-based protocol. A cluster of nodes work together to offer efficiency routing, and the size of each cluster can vary between an upper bound (*ALPHA*) and a lower bound (*BETA*). Each node maintaining O(log(n)) logical links. ABC can achieve each query within $O(\frac{\log(n)}{\log.\log.n})$ hops in the structured P2P system, where *n* is the total number of nodes in the system. CISS [8] is a collaborative information shared system based DHT, it uses a locality based preserving function (LPF) instead of a hash function. CISS consists of a client and server modules. The client module takes the updates of queries, it routes them to rendezvous peer nodes for processing. The server module stores objects to its repository and processes incoming queries, it returns matched results to requesting peer nodes. The key idea of RPS [18] is to partition the key space (*or the identifier space*) into multiple non-overlapping search regions and assign a region to a node that receives a query message. Each node, on receiving a query message, is allowed to forward the query to other nodes only within its search region. Furthermore, at each step of query forwarding, the forwarding node partitions its search region into smaller regions, i.e., the search region starts as the whole identifier space and gets partitioned recursively as the query messages are forwarded. The search region information is carried in the query message as a tag. Pastry [4] assumes a circular identifier space and each has a list connecting of $\frac{L}{2}$ successors and $\frac{L}{2}$ predecessors known as a leat set. A node also keeps track of *M* nodes that are close according to another metric other than the *id* space like network delay. This set is known as the neighbourhood set and is not using during routing but using for maintaining locality properties. The third type of node state is the main routing table. It contains $log_2^b(N)$ rows and $2^b$-1 columns. *L, M* and *b* are system parameters. Tapestry **[1]** is a self-organizing routing and object location system. Like pastry it is based on the earlier work of Plaxton. It provides routing of messages directly to the "closest" copy of an object (*or service*) using only point-to-point links between nodes and without centralized resources. Location information is distributed within the routing infrastructure and is used for incrementally forwarding messages from point to point until they reach their destination. This information is repairable "*soft-state*", its consistency is checked on the fly, and if lost due to failures or destroyed, it is easily rebuilt or refreshed. GTapestry **[13]** assembles physically neighbouring nodes in the Internet into self-organized groups. The routing mechanism of GTapestry is divided into inter-group routing, used by groups that communicate with others through their leaders, and intra-group routing, used by group members to communicate with each other directly.

Kademlia network **[15]** partitions the identifier space exactly like pastry. However, the node *ids* are leafs of a binary tree where each node's position is determined by the shortest unique prefix of its *id*. Each node divides the binary tree into a series of a successively lower subtree that don't contain the node *id* and keeps at least one contact in each of those subtree. Kademilia does not keep a list of nodes close in the identifier space like the leat set or the successor list in Chord. However, for every subtree/interval in the identifier space, it keeps *k* contacts rather than one contact if possible, and calls a group of no more than *k* contacts in a subtree. P4L **[20]** uses a hierarchical rings for content distribution, each ring is similar to Chord in routing, management. Two neighbouring rings communicate between them using a node belong to these two rings (*relay node*). The cost lookup in P4L is $O(\sum_{i=1}^{4} N_i)$ where *ni* is the number of nodes on ring level *i* (*with maximum of 256 nodes in each ring*). Palma **[16]** is location management in mobile environment based on the Tapestry algorithm. Palma architecture is composed of heterogeneous wireless and wired networks connected via a high speed wired backbone network and extended with a number of distributed location servers (LSs). Their LSs are organized into an overlay network to publish location information to each other for storage, and to collaboratively resolve queries. In DPMS **[11]**, advertised patterns are replicated and aggregated by the peers, organized in a lattice like hierarchy. Replication improves availability and resilience to peer failure, and aggregation reduces storage overhead. In DPMS a peer can act as a leaf peer or indexing peer. A leaf peer resides at the bottom level of the indexing hierarchy and advertises its indices (*created from the objects it is willing to share*) to other peers in the system. An indexing peer, on the other hand, stores indices from other peers (*leaf peers or indexing peers*). A peer can join different levels of the indexing hierarchy and can simultaneously act in both the roles. Indexing peers get arranged into a lattice like hierarchy and disseminate index information using repeated aggregation and replication. DPMS uses replication trees for disseminating patterns from leaf peers to a large number of indexing peers. However, such a replication strategy would generate a large volume of advertisement traffic. To overcome this shortcoming, DPMS combines replication with lossy-aggregation, advertisements from different peers are aggregated and propagated to peers in the next level along the aggregation tree. DCFLA **[10]** is a distributed user profile management scheme using distributed hash table (DHT) based routing protocols. SkipNet **[35]** is a ring approach based on the SkipList. This last is a sorted linked list that contains supplementary pointers at some nodes that facilitate large jump in the list in order to reduce the search time of a node in the list. The idea is applied to the ring structure, where nodes maintain supplementary pointers in the circle identifiers space. SkipNet facilitates placement of keys based on nameID scheme that allows the key to be stored locally or within a confined administrative domain (*path locality*). In addition, SkipNet also provides path locality by restricting the lookups in the DHT only to domains that may contain the required key. Koorde **[33]** implements De Bruijn graphs on top of ring architecture. A De Bruijn graph maintains two pointers to each node in the graph, thereby requiring only constant state by node in the ring, specifically, given that each node ID is represented as a set of binary digits, each node is connected to nodes with identifiers 2*m* and 2*m* + 1 (*where m is a decimal value of the node's ID*). These operations can be regarded as simple left shift and additions of the given node's ID. Therefore, by succession shifting of bits, lookup time of *log (n)* can be maintained. Panache **[17]** aggregates popularity information and builds upon other peer-to-peer systems that distribute index information by keyword. Relying on a combination of Bloom filtering, query ordering, and truncated results based on popularity data. Tarzan **[19]** is a P2P anonymizing network layer. A message initiator chooses a path of peers pseudo-randomly through a restricted topology in a way that adversaries cannot easily influence. Tarzan provides anonymity to both clients and server, without requiring that both participate, it uses NAT to bridge between Tarzan host and obvious internet hosts. Cycloid **[9]** is a constant-degree

P2P architecture, which emulates a cube-connected cycles graph in the routing of lookup requests. Cycloid combines Pastry with cube-connected cycle graphs. In a Cycloid system with $n = d*2^d$ nodes at most, each lookup takes $O(d)$ hops with $O(1)$ neighbours per node. Cycloid is not necessarily complete; it can have nodes less than $d*2^d$ with some void node places. Like Pastry, it employs consistent hashing to map keys to nodes. A node and a key have identifiers that are uniformly distributed in a $d*2d$ identifier space. Viceroy **[34]** is based on the butterfly graph, like many other systems, it organizes nodes into a circular identifier space and each node has successors and predecessors pointers. Moreover, in N-nodes network, nodes are arranged in $log_2(N)$ level numbered from 1 to $log_2(N)$. Each node apart from nodes at level 1 has "up" pointer and every node apart from the nodes at the last level 2 "down" pointers. There is one short and one long "down" pointers. Those three pointers are called the butterfly pointers. All nodes also have pointers to successors and predecessors pointers on the same level. In such way, each node has a total of 7 outgoing pointers.

### 2.2.2 Analysis and Discussion

Structured systems offer a scalable solution for exact-match queries, i.e. queries in which the complete identifier of the requested data object is known (*as compared to keyword queries*). There are ways to use exact-match queries as a substrate for keyword queries **[28]**. However, it is not clear how scalable these techniques will be in a distributed environment. The disadvantage of structured systems is that it is hard to maintain the structure required for routing in a very transient node population, in which nodes are joining and leaving at a high rate. Table 2 presents some complexities of structured P2P networks.

**Tab 2.** Complexities of some structured P2P systems

| Architecture | Space complexity | Cost lookup |
|---|---|---|
| Chord | $O(log(n))$ | $O(log(n))$ |
| CAN | $2d$ | $O(n^{1/d})$ |
| Tapestry | $O(log_b(n))$ | $O(log_b(n))$ |
| P4L | $\sum_{i=1}^{4} log(n_i)$ | $\sum_{i=1}^{4} log(n_i)$ |
| Pastry | $O(log_2^b(n))$ | $O(log_2^b(n))$ |
| Viceroy | 7 | $O(log(n))$ |
| Koorde | 2 | $O(log(n))$ |
| Kademlia | $O(log(n))$ | $O(log(n))$ |

## 3. Peer to Peer Network Applications

Since the apparition of P2P network, applications are in a continuously grow, from file sharing to real time applications.

**File Sharing:** content storage and exchange is one of the areas where P2P technology has been most successful. File sharing applications focus on storing and retrieving information from various peers in the network. One of the best known examples of such P2P systems is Emule, KaZAa **[26]**.

**Distributed Computing:** these applications use resources from a member of network computers. The general idea behind these applications is that idle cycles from any computer connected to the network can be used for solving the problem of the other computers that require extra computation. SETI@home **[25]** is one example of such systems.

**Communication and Collaboration:** collaborative P2P applications aim to allow application level collaboration between users. These applications range from instant messaging and chat, to online games, to shared applications that can be used in business, education and home environments. Groove **[36]** and Jabber **[37]** are two examples of such systems.

### Analysis and Discussion

P2P networks are known as a file sharing applications. However, it has several kinds of applications as mentioned above. Each C/S application has emerged to P2P application. P2P networking is not restricted to technology, but covers also social processes with a peer-to-peer dynamic. In such context, social P2P processes are currently emerging throughout society.

## 4. Peer to Peer Network Classifications

According to the degree of decentralization of P2P systems, they are divided on three classes represented on table 3.

**Tab 3.** Degree of decentralization based classification

| Degree of decentralization | Examples |
|---|---|
| Purely decentralized | Gnutella, DHT based architectures |
| Partially centralized | Kazaa, Morpheus |
| Hybrid decentralized | Napster |

In **[21]**, the authors propose a classification based on the type of application. Table 4 presents illustrate this classification.

**Tab 4.** Application categories based classification

| Applications | Examples |
|---|---|
| Communications and collaboration | [37] |
| Distributed computation | [25], [55] |
| Internet Service support | [56], [57] |
| Database systems | [58], [59] |
| Content distribution | [60] |

In **[44]**, the authors propose a classification based on the particularity of each P2P architectures. Table 5 illustrates this classification.

**Tab 5.** Particularity and resemblance based classification

| Taxonomy | Selected references |
|---|---|
| Search | **[40], [41], [43]** |
| Ring | **[5], [20]** |
| De Bruijn Graph | **[47]** |
| Skip Graph | **[45], [46]** |
| Key Words Lookup | **[2], [42]** |
| locality | **[52], [53]** |

P2P networks can be also classified on **1)** semantic consideration on rooting: network with semantic routing and without semantic routing. The former is generally based on flooding or local/distributed index and uses key works as in Gnutella. The later is generally based on DHTs such as Chord, P4l. **2)** physical proximity consideration: for real time application, lookup is measured by time not by the hops number. Most of P2P networks don't consider physical proximity. **3)** Generation according to lifetime cycle: the first generation such as Napster system, the second generation such as Gnutella, and the third generation such as DHT based networks,

**Analysis and Discussion**
Many classifications of P2P networks have been presented on the literature. Generally they are focused on routing strategy, because it is the most important and the most executed operation.

## 5. Peer to Peer Network Security

Distributed implementations create additional challenges for security compared to client-server architecture, security in P2P system aims to ensure that the use of the system does not have unwanted influence to a user or environment where the P2P system operates. Achievement a high level of security in peer-to-peer systems is more difficult than non-peer-to-peer systems **[38]**. The most important attack types are: **1)** Replay Attacks: using a previously recorded or captured message to attack a network or to gain access to somewhere one is not authorized to be (*a form of identity theft*), **2)** Malicious Provider: a provider that accepts payment but fails to complete the transaction can be contested, **3)** Malicious Consumer: a malicious consumer who fraudulently claims that he did not receive services even though he did is thwarted by the use of certificates. The provider simply provides the certificate to his bank-set when the transaction is complete, **4)** Routing Attacks: In such case, message routing will fail with high probability, and the systems fail to provide any services, **5)** Denial of Service Attack**:** a DoS attack is an attempt to prevent legitimate users of a service or network resource from accessing that service or resource, **7)** Sybil Attacks: in a peer-to-peer domain without external identifiers, any node can manufacture any number of identities.

**Analysis and Discussion**
Research concerning security and trust in P2P systems draws upon the expertise of the distributed computing community as well as the sociology community. Even the best protected organizations, companies or personal users are finding it difficult to effectively shield themselves against all malicious security attacks due the increasing rate with which they appear and spread. Distributed implementations of P2P networks create additional challenges for security compared to client-server architecture, especially for reliability, flexibility and load balancing.

## 6. Peer to Peer Platforms

P2P platforms provide infrastructure to support distributed applications using p2p mechanisms. P2P components used in this context are for instance naming, discovery, communication, security and resource aggregation.

**XtremWeb:** it is a P2P project intended to distribute applications over dynamic resources according to their availability and implements its own security and fault tolerance policies **[62]**. XtremWeb manages tasks following the coordinator worker paradigm. The coordinator masters the tasks management process. Workers are distributed volunteer entities which use a part of their CPU time to compute tasks provided by the coordinator. Every worker connection is registered by the coordinator, and it requests task to compute accordingly to its own local policy. The workers download task software and all expected objects, stores them and starts computing. When a task is completed, the worker sends the result back to the coordinator.

**Proactive:** it is a project of ObjectWeb Consortium (*ObjectWeb is an international consortium fostering the development of open-source middleware for cutting-edge applications: e-business, clustering, grid computing, management services ...*) **[63]**. Proactive is a Java library for parallel, distributed and concurrent computing, also featuring mobility and security in a uniform frame- Work with a reduced set of primitives.

**JXTA:** Project JXTA **[22]** is an open source effort to formulate and implement a set of standard P2P protocols that allow a programmer to build any loosely coupled P2P system. JXTA consists of six protocols that support core P2P operations, such as peer discovery, organization, identification and messaging. JXTA architecture is divided into three layers where it implements the OSI model:
**1) Applications Layer:** this layer implements applications that are integrated to JXTA. Many applications are included such as P2P instant messaging and file sharing. JXTA applications implement the OSI application layer. **2) Services Layer:** this layer implements services such as searching and indexing, file sharing, protocol translation, authentication and Public Key Infrastructure (PKI) services, as well as many others. JXTA services implement session, presentation and application layers in the OSI model. **3) Platform Layer (*JXTA Core*):** this layer implements a minimal set of primitives that are common to P2P networking. Primitives include discovery, transport, creation of peers and peer groups and others. The JXTA core implements transport, network and data link layers in the OSI model.

JXTA defines a series of protocols, and XML message formats, for communication between peers **[23]**. Peers use these protocols to advertise and discover network resources, discover each other, and to communicate and route messages. There are six JXTA protocols: Peer Discovery Protocol (PDP), Peer Resolver Protocol (PRP), Peer Information Protocol (PIP), Peer Membership Protocol (PMP), Pipe Binding Protocol (PBP) and Endpoint Routing Protocol (ERP).

**Analysis and Discussion**
The existing P2P networks don't collaborate together even for the same tasks, each one uses its own lookup mechanism. P2P platforms are then introduced; their main objective is to unify the user communities, and then enables for example Gnutella users to search on Freenet network. P2P platforms provide infrastructure to support distributed applications using p2p mechanisms.

## 7. Peer to Peer Network Simulators

Many P2P network simulators have been proposed in the literature these last years. The most important are resumed on **[64]**. DHTSim is a discrete event simulator for structured overlays, specifically DHTs. It is intended as a basis for teaching the implementation of DHT protocols, and as such it does not include much functionality for extracting statistics. P2PSim is a discrete event packet level simulator that can simulate structured overlays only. It contains implementations of six candidate protocols: Chord, Accordion, Koorde, Kelips, Tapestry and Kademlia. OverlayWeaver provides functionality for simulating structured overlays only and does not provide any simulation of the underlying network. It is packaged with implementations of Chord, Kademlia, Pastry, Tapestry and Koorde.
PlanetSim[1] it is an event-based P2P simulator written in Java, it an object oriented simulation framework for overlay networks and services. PeerSim is designed specifically for epidemic protocols with very high scalability and support for dynamicity. It can be used to simulate both structured and unstructured overlays. GPS is a message level discrete event simulator with a built-in protocol implementation of BitTorrent. It allows for simulation of both structured and unstructured overlays.
Neurogridis a P2P search protocol project that includes a single threaded discrete event simulator, originally designed for comparing the Neurogrid protocol, Freenet and Gnutella protocols. The simulator works on the overlay layer level and can simulate either structured or unstructured protocols. It is packaged with implementations of Gnutella, Freenet and the Neurogrid protocols. It is a single threaded discrete event simulator and it does not simulate the underlying network. Query-Cycle Simulator is a P2P file sharing network simulator that uses the Query- Cycle model. In this model, peers, both good and malicious, form an unstructured P2P network. Narses is a scalable, discrete event, flow based application-level network simulator. It allows for modelling of the network with different levels of accuracy and speed to efficiently simulate large distributed applications.

**Analysis and Discussion**
The existing network simulators such as OPNET or NS-2[2] are already used for new P2P models performance evaluation. However, users are confronted by many difficulties, especially on code writing, because these simulators are not conceived for P2P. Many specialized P2P network simulators are then appeared; they integrated more and more routing protocols for facilitating simulation of new P2P models.

## 8. Conclusion and Perspectives

The limitations of client/server systems become evident in large scale distributed environments. P2P networks can be used for improving communication process, optimizing resources discovery/localization, facilitating distributed information exchange. Peer-to- Peer applications need to discover and locate efficiently the node that provides the requested and targeted service. Many P2P architectures have been proposed in the literature, these architectures do not collaborate together, and then P2P platforms have been appeared. Security is an important issue on P2P networking; several attacks are discovered these last years and the major solutions are inspired from these of wireless networks. The existing simulators such as OPNET or NS-2 are not well adapted to P2P. An important number of specialized simulators are then proposed; generally they are focused on routing performance evaluation. This paper presents a generalized and complete survey on P2P activities. Important features that should be addressed on P2P network are performance, scalability, maintenance, reliability, usability, naming, structuring, routing and locating, resource managing, topology updating. As a future works, we envision classifying the mathematical modelling of P2P networks.

---
[1] http://projects-deim.urv.cat/trac/planetsim/wiki/PlanetSim
[2] http://www.isi.edu/nsnam/ns/

### Acknowledgments

The authors would like to thank Mr L. Khenous and Mrs N. Halfoune from Bejaia University for their comments and suggestions.

**Mourad Amad** received the engineer degree from the National Institute of Computer Science (*INI-Algeria*) in 2003 and the magister degree from the University of Bejaia (*Algeria*) in 2005. Currently, he is a PhD at the University of Bejaia, Member of laboratory L.A.M.O.S. His research interests include peer to peer networks (*architecture, application, security, VoIP*)

**Ahmed Meddahi** is a member of GET/Telecom Lille I Computer Science and Networks department. He obtained his Master degree from University of Lille (*France*) and Ph.D. from University of EVRY (*France*) and "Institut National des Telecommunications". His main interests are focused on IP signalling performance,

"VoIP" quality and "context aware" management, P2P. He is associate member of RS2M research group at INT.

**Professor Djamil Assani** was born in 1956 in Biarritz (*Basque Country, France*). He started his career at the University of Constantine in 1978. He received his Ph.D in 1983 from Kiev State University (*Soviet Union*). He is at the University of Bejaia since its opening in 1983/1984. Director of Research, Head of the Faculty of Science and Engineering Science (1999 - 2000). Director of the LAMOS Laboratory (*Modelling and Optimisation of Systems - http://www.lamos.org*), Scientific Head of the Doctoral Computer School (*since its opening in 2003*), he has taught at several universities (Algiers, Annaba, Rouen, Dijon, Montpellier, Tizi Ouzou, Stif,...). He has published many papers on Markov chains, queueing systems, reliability theory, inventory, risk theory, performance evaluation and their applications in some industrial areas as electrical networks and computer systems. He was the president of the national Mathematical Committee (Algerian Ministry of Higher Education and Scientific Research - 1995 - 2005).